\def\lesssim{{_ <\atop{^\sim}}}

\def\ap3m{AP$^3$M}
\def\LCDM{$\Lambda$CDM}
\def\LWDM{$\Lambda$WDM}
\def\hkpc{$h^{-1}{\ }{\rm kpc}$}
\def\hMpc{$h^{-1}{\ }{\rm Mpc}$}
\def\hMsun{$h^{-1}{\ }{\rm M_{\odot}}$}

\def\nbody{$N$-body}
\def\c15{$c_{\rm 1/5}$}

\newcommand{\Fig}[1]{Fig.~\ref{#1}}

\def\ea{et~al.~}                            

\def\lesssim{\mathrel{\hbox{\rlap{\hbox{\lower4pt\hbox{$\sim$}}}\hbox{$<$}}}}
\def\gtrsim{\mathrel{\hbox{\rlap{\hbox{\lower4pt\hbox{$\sim$}}}\hbox{$>$}}}}

\newcommand{\ApJ}[3]    {\mbox{ApJ~\textbf{#1},~#2~(#3)}}

\newcommand{\MNRAS}[3]  {\mbox{MNRAS~\textbf{#1},~#2~(#3)}}
\newcommand{\Nature}[3] {\mbox{Nature~\textbf{#1},~#2~(#3)}}

\newcommand{\PhRevD}[3] {\mbox{Phys.~Rev.~\textbf{D#1},~#2~(#3)}}
\newcommand{\astroph}[1]{\mbox{\texttt{astro-ph/#1}}}
\newcommand{\hepph}[1]  {\mbox{\texttt{hep-ph/#1}}}

\documentstyle[epsfig]{mn}

\begin{document}

\title{Warm Dark Matter versus Bumpy Power Spectra}

\author[Little B., Knebe A., Islam R.]
       {Brett Little$^1$, Alexander Knebe$^1$, Ranty R. Islam$^2$\\        
       {$^1$Centre for Astrophysics \& Supercomputing,
        Swinburne University, P.O. Box 218, Mail \# 31,
        Hawthorn, Victoria, 3122, Australia}\\
       {$^2$Astrophysics, Keble Road, Oxford, OX1 3RH, UK}}

\date{Received ...; accepted ...}

\maketitle

\begin{abstract}
In this paper we are exploring the differences between a Warm Dark
Matter model and a CDM model where the power on a certain scale is
reduced by introducing a narrow negative feature (''dip''). This dip
is placed in a way so as to mimic the loss of power in the WDM model:
both models have the same integrated power out to the scale where the
power of the Dip model rises to the level of the unperturbed CDM
spectrum again.

Using \nbody\ simulations we show that some of the large-scale
clustering patterns of this new model follow more closely the usual
CDM scenario while simultaneously suppressing small scale structures
(within galactic halos) even more efficiently than WDM. The analysis
in the paper shows that the new Dip model appears to be a viable
alternative to WDM but it is based on different physics. Where WDM
requires the introduction of a new particle species the Dip model is
based on a non-standard inflationary period. If we are looking for an
alternative to the currently challenged standard \LCDM\ structure
formation scenario, neither the \LWDM\ nor the new Dip model can be
ruled out based on the analysis presented in this paper. They both
make very similar predictions and the degeneracy between them can only
be broken with observations yet to come.

\end{abstract}


\section{Introduction}
The so-called Cold Dark Matter crisis has led to a vast number of
publications trying to solve the problems which seem to be associated
with an excess of power on small scales. One possibility to reduce
this power is to introduce Warm Dark Matter (i.e. Knebe~\ea 2002;
Bode, Ostriker~\& Turok 2001; Avila-Reese~\ea 2001; Colin~\ea 2000).
But another way to decrease power on a certain scale is to introduce a
negative feature (``dip'') into an otherwise unperturbed CDM power
spectrum.

Several mechanisms have been proposed that could generate such
features in the primordial spectrum during the epoch of
inflation. Among these are models with Broken Scale Invariance (BSI)
(Lesgourgues, Polarski~\& Starobinsky 1998), and particularly BSI due
to phase transitions during inflation (Barriga~\ea 2000). Other
inflationary models include resonant production of particles
(Chung~\ea 1999) and non-vacuum initial states as the quantum
mechanical origin for inflationary perturbations (Martin, Riazuelo \&
Sakellariadou 2000). Broad features can also be introduced into the
power spectrum by including a running spectral index in slow-roll
inflationary models (e.g. Hannestad, Hansen~\& Villante 2000).
However, the parameter range for such broad features is relatively
tightly constrained by observational data from the BOOMERanG (de
Bernardis~\ea 2000) and MAXIMA experiments (Balbi~\ea 2000), and some
of the features tend to be too broad to be able to yield the
scale-dependent effects on the power spectrum we consider here.  After
inflation, other effects such as pressure-induced
oscillations\footnote{This is also recorded in the oscillatory nature
of the spectrum of the cosmic microwave background fluctuations.} in
the matter radiation fluid before decoupling also leave an imprint in
the overall matter transfer function towards small scales mainly in
the form of successive crests and troughs (Eisenstein \& Hu
1998). These features will therefore also be part of the cosmological
matter power spectrum after matter-radiation equality. Only for
extraordinarily large baryon fractions will these features be
significant enough to affect the subsequent formation of small and
medium scale structures.

Most of the above mechanisms appear capable of producing localized
features on any given scale, but to date only large scales have been
tested quantitatively.  The non-vacuum initial state model by Martin,
Riazuelo~\& Sakellariadou (2000), for instance, can lead to a
primordial $P(k)$ with a sharply localized peak, where the location
$k_0$ of that peak is determined by a characteristic built-in scale of
their model. On the other hand, a negative feature is possible BSI
scenarios which lead to oscillations beyond a certain scale in $P(k)$
(Lesgourgues, Polarski~\& Starobinsky 1998). As \nbody\ simulations
are limited in the $k$-range modeled, one might only be able to trace
the first dip of those oscillations due to a privileged energy within
the inflaton potential that determines the scale $k_0$.

Our objective now is to examine the effect of a negative feature in
the primordial power spectrum at very small scales that directly
affects the formation and evolution of (sub-)galactic halos. In
particular we have chosen a negative Gaussian feature on a scale ($k
\simeq 1.8 h{\rm Mpc}^{-1}$). We have not tied our analysis to any of 
the above models, but we have adopted a more generic approach modeling
the dip as a Gaussian in log-space (see below). This differs from
earlier studies of 'bumpy power spectra' (e.g. Lewin~\& Albrecht,
2001). Such a modification is more similar to the idea of Warm Dark
Matter, but the difference lies in the rise of power towards even
smaller scales.  Where WDM cuts off all power exponentially below a
certain threshold, we introduce here a model that has a rather narrow
dip in $P(k)$ and the detailed shape for the loss of power is
completely different, respectively.  We have shown in an earlier paper
(Knebe, Islam~\& Silk 2001) that such localized features in the matter
power spectrum can obscure the interpretation of the evolution of
cluster abundance as an indicator of the cosmological density
parameter. In this paper we are now going to show that a similar
feature (but on smaller scales) can lead to comparable results as for
instance a Warm Dark Matter model. This Dip model can hence be
understood as a plausible alternative to WDM without making
assumptions about the nature of dark matter and only further studies
(and observations) might break the degeneracy between those models
found in this analysis.

The structure of this paper closely follows that of the WDM
investigation by Knebe~\ea (2002). We chose to study the same
properties, but this time focusing on the comparison between the
\LWDM\ and the new Dip model.  We first
start with a brief description of the newly introduced Dip model in
Section~\ref{DipModel}, then move on to the simulations themselves in
Section~\ref{Nbody}. In Section~\ref{Analysis} we analyze the output
from our numerical simulations highlighting the differences in the
\LWDM\ and the Dip model. We finally summarize our results and
conclude in Section~\ref{Conclusions}.

\section{The Dip-Model} \label{DipModel}
The \LWDM\ and the fiducial \LCDM\ model used in this paper are the
same as those used in Knebe~\ea (2002) with the cosmological
parameters $\Omega_0 = 1/3$, $\lambda_0 = 2/3$, $\sigma_8=0.88$,
$h=2/3$, and $m_{\rm WDM} = 0.5$keV for \LWDM.

For the Dip model we are using the same prescription to introduce a
Gaussian feature into an otherwise unperturbed CDM power spectrum as
outlined in Knebe, Islam~\& Silk (2001) and hence the modified power
spectrum follows the equation:

\begin{equation}
	P_{\rm mod}(k) = P(k) \cdot \left[1 - A \exp\left\{-0.4 \left(\frac{\log{k} -
	\log{k_0}}{\sigma_{\rm mod}}\right)^2\right\}\right]
\end{equation}
\noindent
where $P(k)$ is the unmodified spectrum and the parameters $A,
\sigma_{\rm mod}$, and $k_0$ of the dip are given in Table~\ref{DipParam}. 

\begin{figure}
      \centerline{\psfig{file=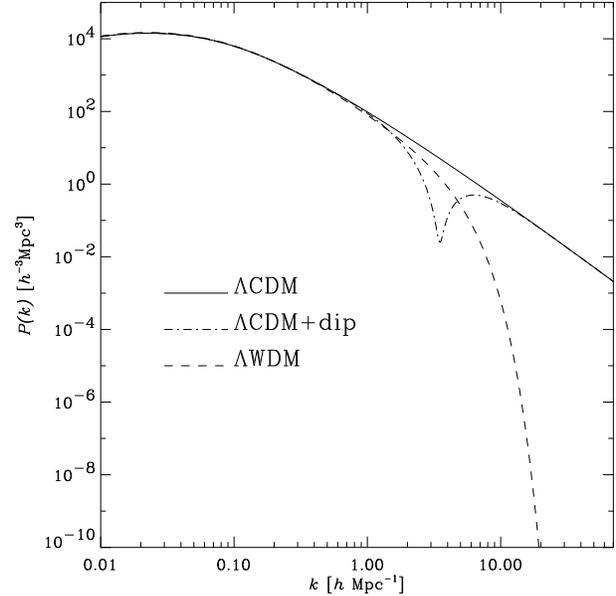,height=\hsize}}
      \caption{CDM input spectra at redshift $z=0$.}
      \label{PkInput}
    \end{figure}

The power taken away from the standard \LCDM\ model via the dip agrees
with the lack of power in the \LWDM\ model to the extend that the
integral

\begin{equation}
 \sigma^2 = {1\over 2\pi^2} \int_0^{k=16h{\rm Mpc}^{-1}} P(k) k^2 dk
\end{equation}

\noindent
is identical for the \LWDM\ and Dip model ($k=16h{\rm Mpc}^{-1}$ is
the point for which the dip meets the unperturbed \LCDM\ power
spectrum again).  The power in the Dip model only drops in a narrow
interval around $k_0$ with the width of that interval controlled by
the parameter $\sigma_{\rm mod}$. A visualization of the three power
spectra used in this study is given in \Fig{PkInput}.

Even though the power that has been taken away from the standard
\LCDM\ model is similar, both models (\LWDM\ and Dip) are based on
different physics. WDM is realized by postulating a finite mass for
the dark matter particles in the range of $m_{\rm WDM} = 0.5 - 2$keV
(0.5keV in our case), whereas the Dip model assumes that the CDM
picture is still valid but based on a primordial power spectrum that
is the result of some non-standard inflationary period as outlined in
the Introduction.

\begin{table}
\caption{Specifications of the artificial modification to the
         \LCDM\ power spectrum.  The mass scale $M = \rho_{\rm crit}
         \Omega \cdot \frac{4\pi}{3}(2\pi/k_0)^3$ corresponding to the
         value of $k_0$ is given in the last column.}
\label{DipParam}

\begin{tabular}{lcccc}\hline
 label     & $2\pi / k_0$ & $A$   & $\sigma_{\rm mod}$ &  mass scale \\ 
\hline \hline
 Dip Model & 1.8\hMpc      & -0.995 & 0.5    &  $2\cdot 10^{12}{\rm M_{\odot}}/h$ \\
\end{tabular}
\end{table}

\section{The $N$-body Simulations} \label{Nbody}
The simulations of the \LWDM\ and the \LCDM\ model were carried out
using the Multiple-Mass ART code (Kravtsov, Klypin~\& Khokhlov 1997)
and the details for these runs can be found in Knebe~\ea (2002). The
new Dip model was set up using the same multiple-mass technique but
evolved with the publicly available Adaptive Mesh Refinement code
\texttt{MLAPM} (Knebe, Green~\& Binney 2001). The number of particles
in use for all three runs was 128$^3$ in a box of comoving side length
25\hMpc.

As both codes (ART and \texttt{MLAPM}) are adaptive mesh refinement
codes designed in a very similar fashion, the technical details
(i.e. refinement criterion, number of integration steps, etc.)  are
identical to the ones already outlined in Knebe~\ea (2002), and hence
we obtained the same force resolution of 3\hkpc\ in the new Dip
model. A quantitative comparison of these two codes can be found in
Knebe, Green~\& Binney (2001) where it is shown that both produce
(nearly) identical results when being run with the same parameter
setup. As the realizations of the initial density field were identical
too for all simulations, we are able to undertake a one-to-one
comparison of the two most massive haloes in all three models.

The halos in each individual model were identified using both the
friends-of-friends algorithm (FOF, Davis~\ea 1985) and the
Bound-Density-Maxima (BDM, Klypin~\& Holtzman 1997) method.

   \begin{figure}
    \centerline{\psfig{file=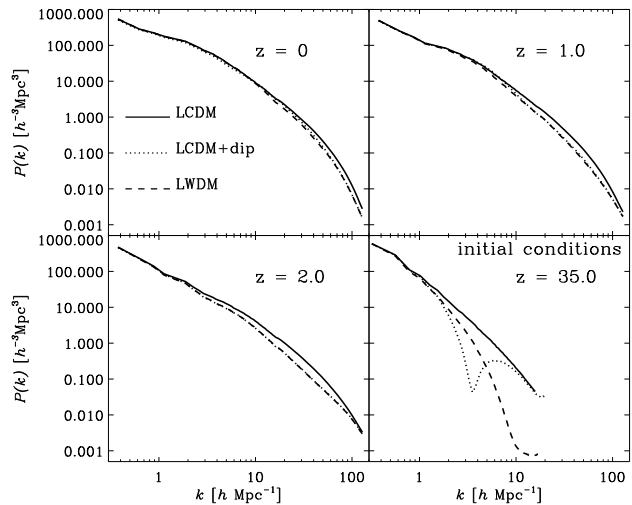,width=\hsize}}
    \caption{Power spectrum evolution.}  
    \label{power}
   \end{figure}

   \begin{figure}
    \centerline{\psfig{file=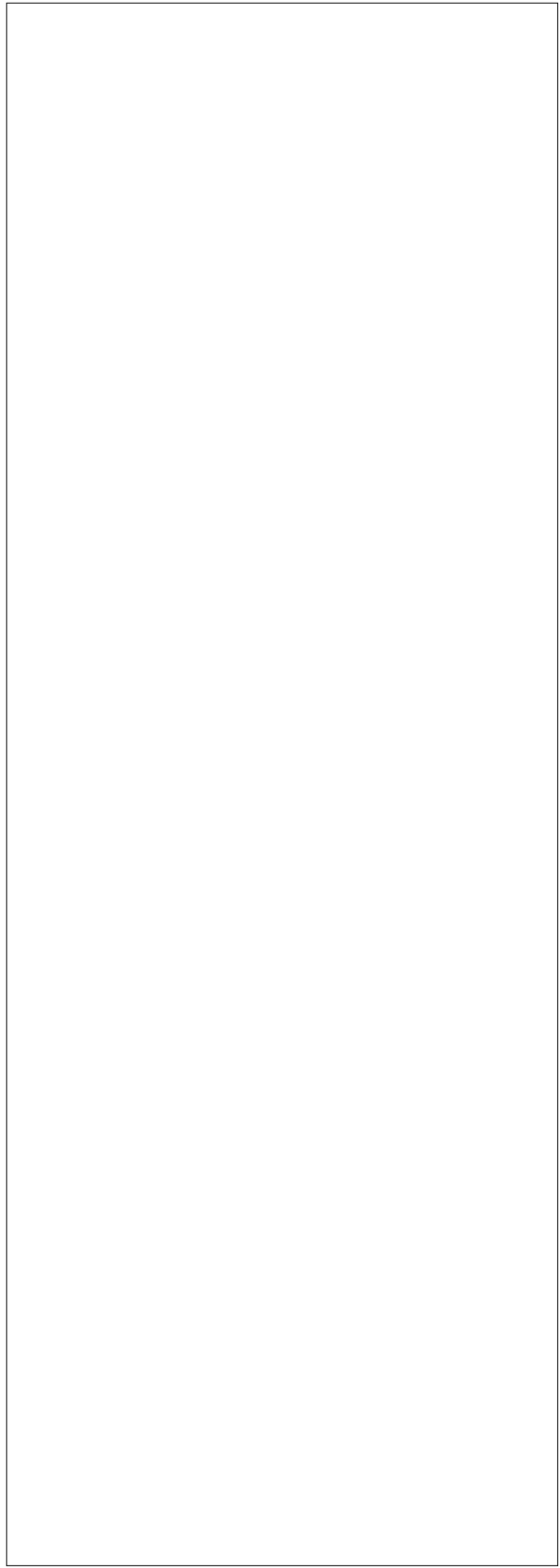,width=8.1cm}}
    \caption{Grey-scaled density field of a 10$\times$10\hMpc\ slice 
             through the simulation at $z=0$.}  
    \label{WDMvsDip}
   \end{figure}

\section{Analysis} \label{Analysis}

\subsection{Power Spectra} 
\label{SECpower}

\Fig{power} shows $P(k)$ as derived from our initial particle distribution
at redshift $z=35$ along with the non-linear evolution at redshifts
$z=2$, $z=1$ and $z=0$.

The Dip is well resolved by our initial conditions and lies inside the
$k$-range defined by the box size $B$=25\hMpc\ ($k_{\rm min} =
2\pi/B$) and the particle Nyquist frequency as determined by the total
number of particles $N$=128$^3$ ($k_{\rm max} = \pi/\Delta x$ with
$\Delta x=B^3/N$).

We can also clearly see that the transfer of power from large to small
scales washes out the features initially present in the power spectrum
at our starting redshift.  The major difference between the WDM and
the Dip model in the simulations is that the loss of power is far
steeper in the latter but also rises again to the level of the
unperturbed \LCDM\ spectrum. In the WDM model the power decreases more
gradually and monotonically.  We will see that this has an effect on
the formation of low-\textit{and} high-mass halos as well as the
amount of substructure to be found in galactic halos. However, the
general predictions of both models remain similar.

Unfortunately we are unable to study the formation process of objects
smaller than 10$^{10}$\hMsun because of the limits imposed by the
total number of particles used. Our mass resolution is ``only" $m_p
\sim 7\cdot 10^8$\hMsun\ and hence we cannot initially resolve the
regime $k>16 h{\rm Mpc}^{-1}$ where the amplitude of the power
spectrum in the Dip and the \LCDM\ models join up again.

    \begin{figure}
    \centerline{\psfig{file=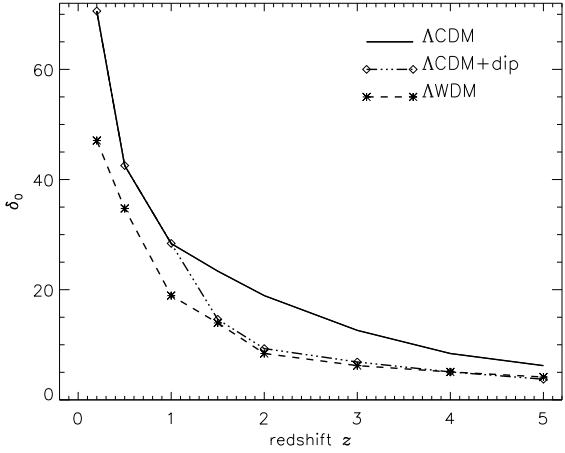,width=\hsize}}
    \caption{Threshold density contrast $\delta_0$ (defined
             via $\dot{\rho}(\delta_0)=0$) as a function
             of redshift for all three models.}  
    \label{MassFlow}
   \end{figure}

\subsection{Dark Matter Density Field} \label{DensityField}
We continue our analysis by inspecting the large-scale structure
density field in all three runs. In \Fig{WDMvsDip} we show a slice
through the particle distribution for all three models with the
particles grey-scaled according to the local overdensity (emphasizing
on low density regions).

We can clearly see the "lumpiness" of \LCDM\ against \LWDM\ and the
Dip model, respectively. But when comparing \LWDM\ with the Dip model
we notice that the filamentary structure of the Universe appears to be
slightly more grainy/evolved in the latter. As pointed out elsewhere
(Knebe~\ea 2002; Bode, Ostriker~\& Turok 2000) small size objects tend
to form in WDM preferably via filament fragmentation as well as at
later times. This phenomenon might not be the case for the Dip model
anymore; we expect to find the filaments to be more clumpy initially
because the power on very small scales rises again to the level of the
"normal" CDM spectrum. And this can be clearly seen in \Fig{WDMvsDip};
the low density dark matter particles appear to be more clustered
(within the filaments) as opposed to a more smooth distribution in
WDM. The similarity between CDM and the Dip model for low-density
regions can be even more emphasized when calculating $\dot{\rho}$, the
flow of matter. To this extent we used the continuity equation

\begin{equation}
\dot{\rho} = \partial \rho / \partial t = - \nabla \cdot (\rho \vec{v}) \ ,
\end{equation}

\noindent
where the rhs was calculated on a 300$^3$ grid covering the whole
computational volume.  The function $\dot{\rho}(\delta)$ with $\delta
= \rho / \overline{\rho} - 1$ has one root in the regime $\delta <
200$ because matter flows out of low density into high density
regions. This defines a 'threshold density' $\delta_0$ via

\begin{equation}
\dot{\rho}(\delta_0) = 0 \ .
\end{equation}

\noindent
The redshift evolution of this $\delta_0$ is presented in
\Fig{MassFlow} for all three models.  We note that $\delta_0$ is at 
all times larger in \LCDM\ than in \LWDM. This can be explained by a 
faster flow of material out of the voids and into the filaments; the
threshold density $\delta_0$ where the inflow of matter equals the
outflow is shifted towards higher density regions. But the Dip model
shows a break at redshift around $z=1.5$. At later times it agrees
with \LCDM\ whereas it matches the \LWDM\ model at higher
redshifts. However, this behavior is difficult to test
observationally as we investigated the flow of low density
material. An analysis of the two-point correlation function of objects
in the mass range \mbox{$10^{10}$\hMsun $< M < 10^{11}$\hMsun} though
showed a higher (and matching) clustering amplitude for \LWDM\ and the
Dip model when compared to \LCDM.

  \begin{figure}
    \centerline{\psfig{file=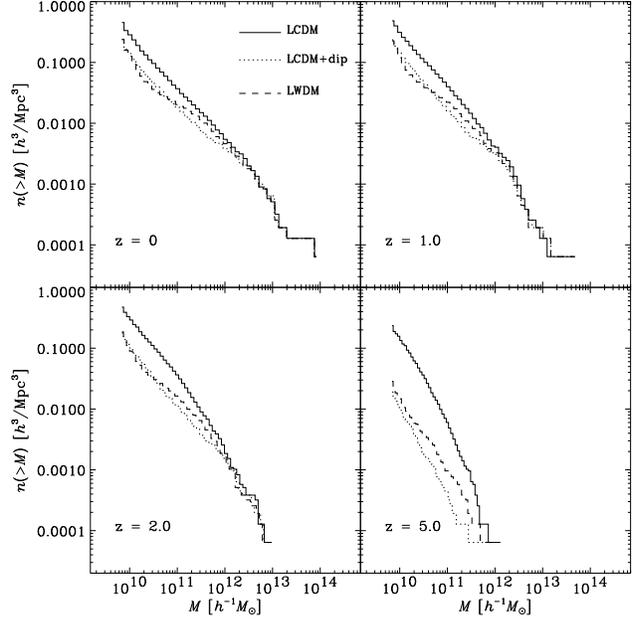,width=\hsize}}
    \caption{Evolution of mass function.}  
    \label{massfunc}
   \end{figure}

\subsection{Mass Function of Halos} \label{mass}
A quantitative difference between the Dip and the \LWDM\ model with
respect to galactic halos becomes apparent when calculating the
(cumulative) mass function~$n(>M)$ as a function of
redshift~$z$. \Fig{massfunc} shows $n(>M)$ for all three models at
redshifts $z=5,2,1$,and $0$. In addition to the fact that the Dip and
the \LWDM\ model show a similar behavior we also note that, first, at
higher redshifts there are fewer
\textit{high} mass objects for the Dip model than for
\LWDM\ and second, the curve for the Dip model for $M<5\times 10^{11}$\hMsun\ 
runs parallel at all times (but lower in amplitude) to the fiducial
\LCDM\ model. 

The former can be explained again by the fact that the drop of power
compared to WDM is sharper in the Dip model; the "excess-lack" of
power is on scales that roughly agree with objects of the order $\sim
10^{11}$\hMsun, which coincides with the high-mass end of mass
function at redshift $z=5$. However, during the course of the
simulation the high mass end of the (Dip-)mass function converges to
the level of the \LCDM\ model.

The latter can be ascribed to the rise of power in the Dip model which
seems to counter act the top-down fragmentation of the filaments found
in the WDM structure formation scenario (i.e. Knebe~\ea 2002; Bode,
Ostriker~\& Turok 2001) in such a way that they tend to be more grainy
initially (as already seen in \Fig{WDMvsDip}). But those small particle 
groups do not make it into larger objects as we lack power on the
"connecting" scales; they stay isolated within the filaments.

\begin{figure}
    \centerline{\psfig{file=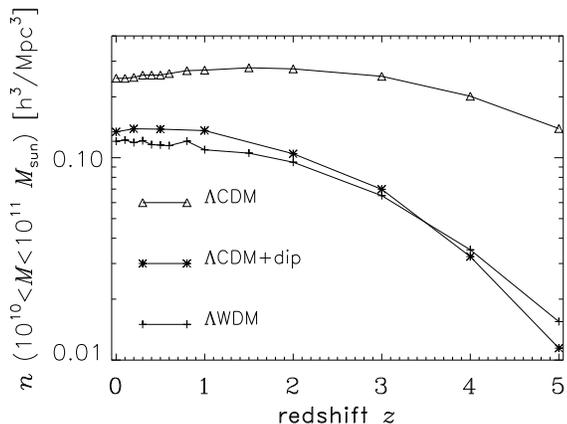,width=\hsize}}
    \caption{Evolution of halo abundance for particle groups with 
             mass $M$ in the range [$10^{10}$\hMsun,$10^{11}$\hMsun].}  
    \label{abundance}
   \end{figure}

In \Fig{abundance} we highlight the number density evolution of
objects in the mass range \mbox{$10^{10}$\hMsun $< M < 10^{11}$\hMsun}
by plotting the abundance evolution for particle groups within that
range out to redshift $z=5$. Again, the behavior for WDM and the Dip
model are very similar, but we see the trend for a steeper and faster
evolution in the latter.

\subsection{Individual Halos} \label{individual}
\begin{figure}
    \centerline{\psfig{file=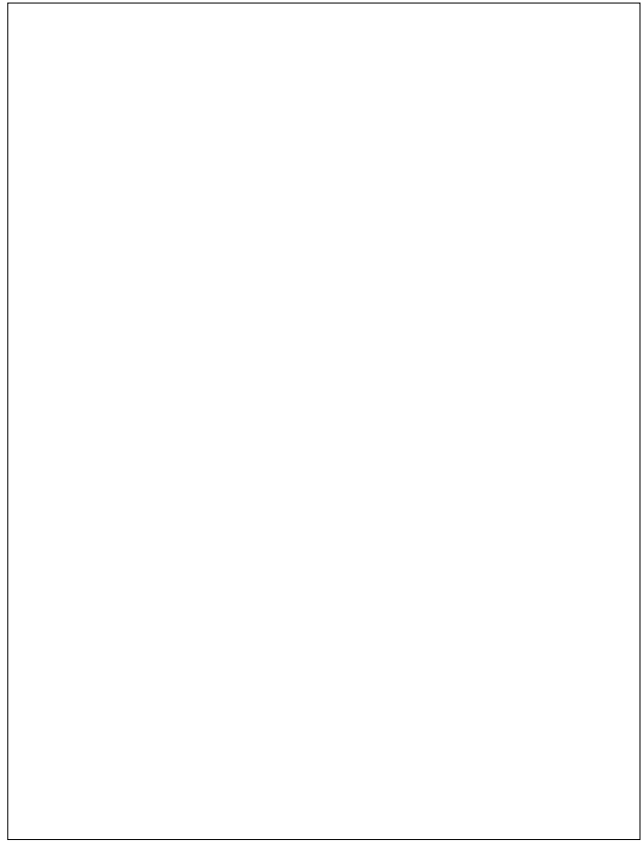,width=\hsize}}
    \caption{Gray-scaled density field of the two most massive
             galactic halos identified at z=0. The left panel shows halo~\#1
             of mass $9.3\cdot 10^{13}$\hMsun\ (about 135000 particles) and
             the right panel shows halo~\#2 weighing $8.2\cdot
             10^{13}$\hMsun\ (ca. 120000 particles).  The upper most panel is
             for the \LCDM\ model, the middle panel shows the \LWDM\ run, and
             the lower panel is the new Dip model.}  
    \label{WowHalos}
   \end{figure}

The remaining analysis will focus on the two most massive halos found
in all three runs.  To this extent we start with showing the density
fields of those two objects. \Fig{WowHalos} confirms again that the
damping of small scale power in the primordial power spectrum leads to
a suppression of substructure in galactic halos. But we can also see
that this suppression is far more pronounced in the Dip model than in
the WDM structure formation scenario. We attribute this again to the
steeper cut-off in the primordial power spectrum.

\begin{figure}
    \centerline{\psfig{file=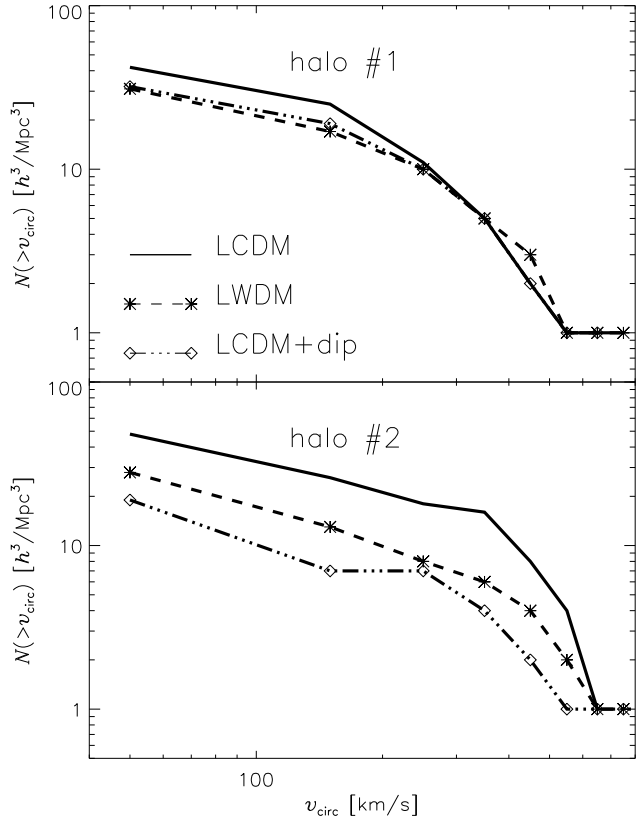,width=\hsize}}
    \caption{Cumulative circular velocity distribution of satellites
             orbiting in halo~\#1 and halo~\#2 for $z=0$.}  
    \label{Nvcirc}
   \end{figure}

To quantify these finding we calculated the cumulative circular
velocity distribution of satellites within halo~\#1 and halo~\#2.  The
result can be found in \Fig{Nvcirc} which is accompanied by
Table~\ref{Nsat} where we give the total numbers of satellites found
in each halo at redshifts $z=0$ and $z=1$, respectively.

\begin{table}
\caption{Number of satellite galaxies orbiting in halo~\#1 and halo~\#2.}
\label{Nsat}
\begin{tabular}{lcccc}\hline
 model     & halo~\#1 & halo~\#2         & halo~\#1 & halo~\#2\\
           & \multicolumn{2}{c}{$z=0$}   & \multicolumn{2}{c}{$z=1$}\\
\hline \hline
 \LCDM     &    42    &  48  &    29    &  12\\
 \LWDM     &    29    &  28  &    29    &  11 \\
 \LCDM+dip &    30    &  19  &    29    &  11 \\
\end{tabular}
\end{table}

There is clear evidence that the Dip model produces even fewer
satellites than WDM. Even though there are roughly the same number of
satellites at redshift $z=1$ in all three models, their number only
marginally increases in the Dip model (especially for halo~\#2). We
previously interpreted the suppression of substructure in WDM by
balancing the accretion and destruction rate of satellites in the halo
(Knebe~\ea 2002); as such, the even higher destruction rate in the Dip
model can best be explained by the presence even more loosely bound
satellites. This again can be ascribed to the faster drop-off in power
and can therefore be controlled by the amplitude $A$ and the width
$\sigma_{\rm mod}$ of the Dip. However, tuning the Dip parameters to
bring the number of satellites into agreement with the WDM model will
automatically have an effect on, for instance, the mass function as
given in
\Fig{massfunc}. The lower the amplitude the closer it should get to
the unperturbed \LCDM\ model.  Hence the Dip model provides a way of
suppressing substructure but otherwise reproducing the standard CDM
structure formation scenario.  The even milder increase in number of
satellites for halo~\#2 in the Dip model (cf. Table~\ref{Nsat} and
\Fig{Nvcirc}) is readily explained by the fact that that particular
halo underwent a major merger near redshift $z=0.2$. The tidal effects
of such a merger tend to destroy the loosely bound satellites in that
model more successfully and hence we are left with even fewer than in
the \LWDM\ model.

\section{Summary and Discussion} \label{Conclusions}
We investigated the differences of two models that both suppress small
scale fluctuations in the primordial power spectrum, namely a warm
dark matter and the newly introduced Dip model.  While it may be
argued that the \LWDM\ model is better constrained in terms of the
allowed masses and thus free streaming length of the potential
particle candidates, the Dip model considered here makes no assumption
about the nature of the dark matter component. This happens at the
expense of introducing two additional parameters (the amplitude $A$
and width $\sigma_{\rm mod}$ of the dip), although these are
determined entirely by inflationary physics and thus before decoupling
(on the basis of the models considered in the Introduction). Using
both the \LWDM\ and the Dip model as the input to \nbody\ simulations
we showed that their predictions are fairly similar with a few notable
differences.

The dip in our model was chosen to mimic the loss of power in a WDM
Universe in a sense that the integrated power (out to the point where
the dip reaches the unperturbed \LCDM\ level again) in both models is
identical. Therefore all differences to be found in the structure
formation process are due to the way in which power on small scales is
suppressed and the detailed shape of the power spectrum, respectively.
The analysis presented in this paper showed that the way the loss of
power is realized is only of minor importance for objects within the
mass range under investigation. However, there are some
quantitative differences, mainly in the shape of the cumulative mass
function at the low mass end and the flow of low density material.

For the Dip model the low density material appears to be more in line
with that expected for CDM (cf. \Fig{MassFlow}). This agreement may
reflect the rise in power to that of CDM on very small scales. To
quantify this effect more accurately, the resolution has to be
improved significantly, but on theoretical grounds this behavior is
expected.  In filaments the `background overdensity' contributed by
large wave length modes tends to be lower than in highly clustered
regions and consequently the onset of non-linear evolution of the
smallest scales will occur later than in cluster environments. In this
case traces of the rise in the linear power spectrum at very small
scales may be more likely to be observable here than in the clustered
regions at the present day.  Observations of structures residing in
filaments could therefore potentially help to constrain the power
spectrum on these very small scales.

Despite the similarities in filamentary structure of the Dip and CDM
model compared to the WDM model, the suppression of low-mass halos is
even more pronounced.  This even stronger suppression of substructure
(in galactic halos) can easily be explained by the amplitude of the
dip and an adequate choice for $A$ will definitely eliminate the
deviation from the WDM model. But in that case one will also
automatically alter the formation times and sites for low mass halos
in general bringing it more into agreement with CDM. The Dip model
hence provides a way of fine tuning the number of satellites without
deviating too far from the standard CDM paradigm (cf.
\Fig{MassFlow} and \Fig{massfunc}) otherwise.

The other difference between WDM and the Dip model is the shape of the
cumulative mass function and the evolution and formation times of
halos. Here we showed that even though small mass objects (as well as
high mass objects) are suppressed, the shape of $n(>M)$ follows more
closely that of the \LCDM\ model; the power law behavior at the low
mass end is preserved. We interpreted this result with respect to the
rise of power in the Dip model which counter acts the top-down
fragmentation of filaments. The filaments are more grainy/evolved
initially due to the relative excess of power on very small scales
compared to the WDM model (cf. \Fig{WDMvsDip}). Moreover, the lack of
high-mass halos at high redshift in the Dip model might allow us to
observationally distinguish between WDM and the Dip model.

The resolution of our simulations did not allow for a more detailed
comparison of the models on very small (mass) scales where the Dip
model has the same power as the \LCDM\ model again. This we postpone
to a future investigation.

To conclude, we remark that the Dip model has proven to be a viable
alternative to the WDM structure formation scenario and only future
observations as well as better resolved numerical simulations will
be able to disentangle the degeneracy between those two models.  The
only way of breaking this degeneracy might lie within the filamentary
structures of the Universe and the low-mass end of the cumulative mass
function. Our results suggested a simple power-law behavior for these
scales (similar to the \LCDM\ model) whereas the shape of the mass function 
in the WDM case is more complicated.  But to strengthen these findings
we clearly need more detailed and much better resolved simulations in
the future.


\section*{Acknowledgments}
We benefitted from valuable discussions with Brad Gibson.  The
simulations presented in this paper were carried out on the Beowulf
cluster at the Centre for Astrophysics~\& Supercomputing, Swinburne
University and the Oxford Supercomputer Centre. We are grateful to
Anatoly Klypin and Andrey Kravtsov for kindly providing a copy of the
ART code that was used for running the \LCDM\ and \LWDM\ simulation,
respectively. We acknowledge the support of the Swinburne University
Research Development Grants Scheme.


\end{document}